\documentstyle[12pt]{article}

 1
 1
 1

\begin{document}

\def\kln{{\kappa_{L}}^{NC}}
\def\krn{{\kappa_{R}}^{NC}}
\def\klc{{\kappa_{L}}^{CC}}
\def\krc{{\kappa_{R}}^{CC}}
\def\ttz{{\mbox {$t$-${t}$-$Z$}\,}}
\def\bbz{{\mbox {$b$-${b}$-$Z$}\,}}
\def\tta{{\mbox {$t$-${t}$-$A$}\,}}
\def\bba{{\mbox {$b$-${b}$-$A$}\,}}
\def\tbw{{\mbox {$t$-${b}$-$W$}\,}}
\def\tltlz{{\mbox {$t_L$-$\overline{t_L}$-$Z$}\,}}
\def\blblz{{\mbox {$b_L$-$\overline{b_L}$-$Z$}\,}}
\def\brbrz{{\mbox {$b_R$-$\overline{b_R}$-$Z$}\,}}
\def\tlblw{{\mbox {$t_L$-$\overline{b_L}$-$W$}\,}}
\def\beq{\begin{equation}}
\def\enq{\end{equation}}
\def\ra{\rightarrow}
\def\ppbar{ {\rm p} \bar{{\rm p}} }
\def\ifb{ {\rm fb}^{-1} }
\def\ETslash{\not{\hbox{\kern-4pt $E_T$}}}
\def\pslash{\not{\hbox{$p$}}}
\def\ggtt{ q \bar q, \, g g \ra t \bar t }
\def\width{ \Gamma( t \ra b W^+) }
\def\ttbar{ t \bar t }
\def\hatt{ \hat {\rm T} }
\def\d{ {\rm{d}} }
\def\doublespaced{\baselineskip=\normalbaselineskip\multiply
    \baselineskip by 150\divide\baselineskip by 100}
\doublespaced
\pagenumbering{arabic}
\begin{titlepage}
\begin{flushright}
\large{
MSUHEP-41009\\
November, 1994}
\end{flushright}
\vspace{0.4cm}
\begin{center} \LARGE
{\bf Strategies for Probing CP Properties \\
in the Top Quark System \\
at $e^-e^+$ and Hadron Colliders }
\end{center}
\begin{center}
{\bf C.--P. Yuan\footnote{Talk presented at The First International
Symposium on ``Symmetries in Subatomic Physics'', May 16-18, 1994,
Taipei, Taiwan, R.O.C.
}}
\end{center}
\begin{center}
{
Department of Physics and Astronomy \\
Michigan State University \\
East Lansing, MI 48824, USA}
\end{center}
\vspace{0.4cm}
\raggedbottom
\setcounter{page}{1}
\relax

\begin{abstract}
\noindent
I discuss strategies for probing CP properties in the top quark system
at $e^-e^+$ and hadron Colliders.
The magnitudes of CP violation effects predicted by various models
are reviewed. I also discuss the potential of various
current and future colliders in measuring the CP asymmetry
associated with the productions and/or decays of the top quarks.
\end{abstract}

\vspace{2.0cm}

\end{titlepage}
\newpage
\section{ Why Top Quarks? }
\indent

For nearly 30 years after the discovery of the CP-violating
decays of the $K^0_L$ meson \cite{kcpx},
the evidence for CP violation remains confined to the neutral kaon system.
The Standard Model (SM)
explains this phenomenon through a phase in the
Kobayashi-Maskawa (KM) matrix \cite{kmcp},
that also predicts CP-violating effects in the $B$ (Beauty) system
which is the main goal of the planned $B$-factories \cite{bfactory}.
It remains true however that all the known CP-violating
phenomena can also be explained by a naive phenomenological model,
called the superweak model \cite{superw}.
It's also possible that other sources of CP violation besides the KM phase
exist in nature.
Various models have been proposed to explain the CP violation effects
in the kaon system and predict new effects in either
the bottom or top quark system. Such models often introduce
new sources of CP violation by
having a richer particle spectrum with particle masses at the weak scale
$v = 246 \,$GeV.
In many of these models,
the CP violation is a consequence of the presence of many
additional couplings
which  mix with the  SM couplings either at the tree level
or at the loop level to induce CP violation.
Some of the studies on CP violation
in the top quark system are given in Ref.~\cite{toppol}-\cite{loopcp}.

For light quarks, an asymmetry in the production of different helicity
states would be unobservable because the polarization of the quark
would likely be washed out when the light quark is hadronized.
The life time ($\tau = 1/\Gamma
\sim (m_t/150)^{-3} \,{\rm GeV}^{-1}$)  of a heavy top quark
($m_t > 100\,$GeV) is shorter than the typical hadronization
time ($\sim 1/\Lambda_{QCD} \sim 1/0.3 \,{\rm GeV}^{-1}$), so the
top quark decays through the
weak interaction before it hadronizes \cite{thad}.
Since the top quark decays through the parity--violating
weak interaction,
$ t \ra b W^+$, its polarization can be self--analyzed
by either the polarization of the $W^+$ or the kinematics of
its decay particles ({\it e.g.}, $b$, $\ell$, and $\nu$)
\cite{toppol}. Because of the large mass of the top quark, it is
sensitive to any new physics which grows as the heavy fermion mass,
such as the interactions mediated by Higgs bosons \cite{thiggs}.
For instance, the Weinberg model \cite{weinberg}
extends the SM Higgs doublet necessary for
symmetry breaking to $n$ Higgs doublets,
when $n \geq 3$ the mass
matrix for the Higgs sector has enough free parameters to allow
complex CP-violating phases to induce CP violation effects
\cite{hzero,hplus}. The neutral Higgs sector is unique in the sense
that a single Higgs boson coupling to a massive fermion is enough
to manifest CP violation as long as the Yukawa coupling contains
both scalar and pseudoscalar components \cite{hzero}.
Thus, due to its large mass, the top quark represents a unique probe
for detecting CP violation.
The Kobayashi-Maskawa mechanism  of  CP violation
in the SM predicts a very small effect for
the top quark \cite{cpsm}. Hence, the top quark system is sensitive to
non-standard sources of CP violation.

As noted in Ref.~\cite{imkane}, it would be difficult to detect a
CP-violating effect less than the order of $10^{-3}$ at hadron colliders
due to the precision of experimental measurements and the
understanding of the intrinsic background processes to the signals
of interest. Here I will report some of the results
in the literature with large CP-violating effects in the top
quark system.
Undoubtedly, a heavy top quark could induce CP-violating effects
in light quark systems, such as in the kaon or the Beauty system.
Here, I will only discuss the final states involving top
quark(s) because only those processes can show large CP violation
effects and could probably be detected at colliders.

\section{ What do we know about the Top Quark? }
\indent

Assuming that the properties
of the top quark interactions are described
by the SM, then
based upon analysis of a broad range of electroweak data, the mass of the SM
top quark is expected to be in the vicinity of 150 to 200 GeV, e.g.,
$169 ^{+16+17}_{-18-20}\,$GeV \cite{mele}.
{}From the direct search at the Tevatron, assuming a SM top quark,
$m_t$ has to be larger than 131\,GeV \cite{D0}.
More recently, data were presented by the CDF group at
FNAL for the evidence of a heavy top quark with mass
$m_t = 174 \pm 10 \, {\rm (stat)} \pm 12 {\rm (syst)}$\,GeV
\cite{CDF}, although the latest analysis of
D0 data did not find a significant signal at high masses \cite{d0top}.

As argued in the previous section, the top quark is sensitive to new
physics because of its large mass. Thus, it is quite possible that,
for instance, the interaction of $\tbw$ is not standard.
Without assuming the top quark to be of the SM nature,
we found that the LEP data do not give useful constraints on the
couplings of $\tbw$,  and $m_t$ can be as large as 300 GeV
depending upon the
strength of the non-standard couplings in $\tbw$ \cite{ehab}.
However, the right-handed charged current of the top quark
is well constrained by the  $b \ra s \gamma$ data
because the magnetic moment form factor requires a spin flip
in its amplitude \cite{fuji}.

In the SM, the coupling of $\tbw$ violates parity and is
purely left-handed at the Born level.
New physics might modify the coupling of the $\tbw$ to induce
CP-violating effects in the production or the decay of the
top quark. The subject of this talk is
to discuss what's the typical size of the CP-violating effect in
top quark system, and how to detect CP-violating effects
at colliders.

\section{ How to Produce the Top Quark? }
\indent

First, let's ask how the top quark is produced at the
current and future colliders.
When giving the production rates, I will assume a
SM top quark.

In $e^-e^+$ collisions, a $\ttbar$ pair is produced via
the electroweak process $e^-e^+ \ra \gamma,\,Z \ra \ttbar$.
A single-$t$ (or single-$\bar t$) event is produced from
$e^-e^+ (W^+ \gamma, \, W^+ Z) \ra  e^- {\bar \nu_e} t {\bar b}$.
In $e^+ \gamma$ collisions, a single-$t$ event
can be produced from $e^+ \gamma \ra {\bar \nu_e} t {\bar b}$.

At the Tevatron (a proton-antiproton collider with $\sqrt{S}=2\,$TeV)
or the Di-TeV (the upgrade of the Tevatron,
a proton-antiproton collider with $\sqrt{S}=4\,$TeV),
the primary production mechanisms for a SM top quark are
the QCD processes $q \bar q, \ gg \ra t \bar t$ .
For a heavy top quark, $m_t > 100$ GeV, the $q \bar q$ process
becomes most important at these energies.
The full next-to-leading-order calculation of these QCD processes was
completed several of years ago \cite{nloqcd}.
The electroweak radiative corrections to these processes were also
calculated in Refs.~\cite{qcdtpol} and \cite{qqttqcd}.
Consequently, the production rates for top quark pairs
at hadron colliders are  well predicted.

If the top quark is as heavy as 175 GeV then another production
mechanism, known as the $W$-gluon fusion process,
$q g (W^+ g) \ra q' t \bar b$,
which produces either a single $t$ or a single $\bar t$ in each
event,
is also important \cite{wgone,wgtwo}.
Eventually, it becomes more important than the QCD processes for
a much heavier top quark, $ m_t \geq 250\,$GeV.
The production mechanism of the
$W$-gluon fusion process involves the electroweak
interaction, therefore it can probe the electroweak sector of
the theory. This is in contrast to the usual QCD production mechanism
which only probes the QCD interaction when counting the
top quark production rates.

At the LHC (a proton-proton collider with
$\sqrt{S}=14\,$TeV) the dominant production mechanism
for a SM top quark is
the QCD process $gg \ra t \bar t$. The subprocess
$q \bar q \ra t \bar t$ is always small compared with the gluon-gluon
fusion process, even when the $t \bar t$ invariant
mass is near the TeV region.
The $W$-gluon fusion process is also important; its production rate
is about an order of magnitude smaller than the $\ttbar$ pair
rates from the QCD processes for a 175 GeV top quark and becomes more
important for a heavier top quark.

\section{ How Does the Top Quark Decay? }
\indent

For a SM top quark, heavier than the $W$--boson, the dominant decay
mode is
the weak two--body decay $t \ra b W^+$.
In this mode, the top quark will analyze its own
polarization \cite{toppol}.
The branching ratio (Br) for the leptonic decay mode
of the top quark, $t \ra b W^+ (\ra \ell^+ \nu_{\ell} )$,
is about $1/9$ for either $\ell= e$, $\mu$, or $\tau$.
The Br of its hadronic decay mode is
about $6/9$ for $t \ra b W^+ (\ra \,{\rm jets}\,)$.

An extension of the standard Higgs sector with two Higgs doublets
has both charged and neutral Higgs bosons. If the charged
Higgs boson is lighter than the top quark, the branching ratio for the decay
$t \ra b H^+$ could be comparable to that for $t \ra b W^+$
\cite{csltwo}.

Another interesting channel for the decay of the top quark is
the flavor changing neutral current (FCNC) decay mode.
In the SM, the branching ratios for the FCNC decay modes
were found to be too small
to be detected: ${\rm Br}(t \ra cH) \sim 10^{-7}$,
${\rm Br}(t \ra cg) \sim 10^{-10}$, ${\rm Br}(t \ra cZ) \sim 10^{-12}$,
${\rm Br}(t \ra c\gamma) \sim 10^{-12}$ \cite{cslfour}.
The branching ratios of these modes
in two Higgs doublet models
or the Minimum Supersymmetric Standard Model (MSSM)
could be enhanced by 3--4 orders of magnitude
if one pushes the parameters far enough \cite{cslfive}.
It is a prediction of the SM and the MSSM that no large FCNC decays
exist for top quarks, so if any are detected they are
 beyond these approaches.
In some models, the branching ratio of the FCNC decay channel
$~t \ra cH~$  may be significantly enhanced, of the order $1\%$,
due to large Yukawa couplings \cite{hou}.

\section{ CP-violating Observables }
\indent

It is known that explicit CP violation requires the presence of both the CP
non-conserving vertex and the complex amplitude.
Due to the origin of this complex structure, the possible CP-violating
observables can be separated into two categories.
In the first category, this complex structure comes from the absorptive
part of amplitude due to the final state interactions.
In the second category, this complex structure does not arise from the
absorptive phase but from the correlations in the kinematics of the
initial and final state particles involved in the physical process.
Hence, it must involve a triple product correlation (i.e., a Levi-Civita
tensor).

To distinguish the symmetry properties between these two cases,
we introduce the transformation $\hatt$, as defined in Ref.~\cite{hatt},
which is simply the application of time reversal to all momenta and
spins without interchanging initial and final states.
The CP-violating observables in the first category are CP-odd and
CP$\hatt$-odd, while those in the second category are
CP-odd and CP$\hatt$-even. Of course, both of them are CPT-even.

As an illustration of the above two categories, we consider the CP-violating
observables for the decay of the top quark.
Consider the partial rate asymmetry
\begin{eqnarray}
{\cal A}_{bW} & \equiv &
    { \Gamma( t \ra b W^+) - \Gamma({\bar t} \ra {\bar b} W^-)
    \over \Gamma( t \ra b W^+) + \Gamma({\bar t} \ra {\bar b} W^-) }.
\end{eqnarray}
This observable clearly violates CP and CP$\hatt$ and therefore belongs to
the first category.
We note that
because of CPT invariance, the total decay width of the top quark
$\Gamma(t)$ has to equal the total decay width of the top anti-quark
$\Gamma(\bar t)$.
Thus, any non-zero ${\cal A}_{bW}$ implies that
there exists a state (or perhaps more than one state) $X$ such that
$t$ can decay into $X$, and ${\bar t}$ into ${\bar X}$.
The absorptive phase
of $t \ra b W^+$ is therefore generated
by re-scattering through state $X$, i.e.,
$t \ra X \ra b W^+$, where $X \neq b W^+$ because
the final state interaction should be off-diagonal \cite{wolf}.

Next, let's consider the observable of the second category.
In the decay of $t \ra b W^+ (\ra \ell^+ \nu_{\ell})$, for a polarized $t$
quark, the time-reversal invariance (T) is violated if
the expectation value of
\begin{eqnarray}
\vec{\sigma}_t \times \vec{p}_b \cdot \vec{p}_{\ell^+}
\end{eqnarray}
is not zero \cite{toppol}. Assuming CPT invariance, this
implies CP is violated.
Therefore, this observable is CP-odd but CP$\hatt$-even.
A non-vanishing triple product observable,
such as Eq.~(2), from the decay of the top quark
violates T, however it may be entirely due to final state interaction
effects without involving any CP-violating vertex.
To construct a truly CP-violating observable, one must combine
information from both the $t$ and $\bar t$ quarks.
For instance, the difference in the expectation values of
$\vec{\sigma}_t \times \vec{p}_b \cdot \vec{p}_{\ell^+}$
and
$\vec{\sigma}_{\bar t} \times \vec{p}_{\bar b} \cdot
\vec{p}_{\ell^-}$
would be a true measure of an intrinsic CP violation.

If the polarization of the $\tau$ lepton in the
decay of $t \ra b \tau \nu_{\tau}$
can be measured, then it has been shown in Ref.~\cite{taupol}
that the CP-violating transverse polarization asymmetry of
the $\tau$ can be of the order of a few tens percent, which
is larger than the typical partial rate asymmetry by about
a factor of 100 ($\sim m_t/m_\tau$). Two kinds of CP-violating polarization
asymmetries can then be constructed. One falls into the first category,
another into the secondary category. We refer the reader to
Ref.~\cite{taupol} for more details.  Experimentally, this would be
a big challenge because one needs to determine the moving directions of
both the $t$ and $b$ quarks
and the polarization of the $\tau$ to measure such asymmetries.

\section{ CP violation in Top Pair Productions }
\indent

Many studies have been done in the literature on how to measure the
CP-violating effects in the top quark system~\cite{toppol}-\cite{loopcp}.
Some of the results for $e^-e^+$ and hadron colliders
are summarized in the following.

\subsection{ At $e^-e^+$ Colliders }
\indent

In $e^-e^+ \ra t \bar t$, the modes of $t_L {\bar t}_R$ and
 $t_R {\bar t}_L$ are self-conjugate, but  $t_L {\bar t}_L$
 and $t_R {\bar t}_R$ are CP conjugate of each other. Therefore, the
difference between the event rates $N(t_L {\bar t}_L)$ and
$N(t_R {\bar t}_R)$ signals a CP asymmetry.
(We adopt the notation that $t_L$ is a left-handed top quark,
and ${\bar t}_L$ is a left-handed top anti-quark.
$N(t_L {\bar t}_L)$ denotes the number of events with a left-handed
$t$ and $\bar t$ pair.)

If we assume that new physics only comes in the production mechanism
of $t \bar t$, then the asymmetry in
$N(t_L {\bar t}_L) - N(t_R {\bar t}_R)$
can be measured from the energy asymmetry in
the leptons~\cite{hzero,eecp,hatt}, which is sensitive only to the
absorptive parts of CP-violating form factors.
The CP-violating asymmetry
\begin{eqnarray}
{\cal A}_E(\ell) &  \equiv &
{ {\d \sigma \over \d E(\ell^+) } -  {\d \sigma \over \d E(\ell^-) }
\over {\d \sigma \over \d E(\ell^+) } + {\d \sigma \over \d E(\ell^-) }
}
\end{eqnarray}
therefore belongs to the first category,
where $E(\ell^+)$ is the energy of $\ell^+$ in the center-of-mass (CM)
frame of $t \bar t$.
To measure the asymmetry ${\cal A}_E(\ell)$, all the decay modes
of $t \bar t$ events with either single $\ell$ or double
$\ell$'s can be included to enhance the statistics.
In the CM frame of
$t {\bar t} \ra \ell^+ \ell^- \nu \bar{\nu} b {\bar b}$,
both the $\ell^+$ and $\ell^-$ tend to move along the direction
of $t_R$ (or ${\bar t}_L$) in $t_R {\bar t}_R$ (or $t_L {\bar t}_L$)
events. Thus, if CP is conserved,  there will be equal numbers of
$t_R {\bar t}_R$ and $t_L {\bar t}_L$ produced, and
${\cal A}_E(\ell)$ will be exactly zero.
A non-vanishing ${\cal A}_E(\ell)$ would indicate
CP is violated in $t \bar t$ production.

One of the CP-violating observables from the second category is the
integrated up-down asymmetry ${\cal A}_{ud}$~\cite{hatt}.
Define the $e^-e^+ \ra t \bar t$ scatter plane to be the $x$-$z$ plane.
Let $N(\ell^+,{\rm up})$ denote the number of $t \bar t$ events with $\ell^+$
above the $x$-$z$ plane, i.e., $p_y(\ell^+) > 0$, etc. Then,
\begin{eqnarray}
{\cal A}_{ud} & \equiv & { {  [ N(\ell^+,{\rm up}) +  N(\ell^-,{\rm up}) ]
                - [ N(\ell^+,{\rm down}) + N(\ell^-,{\rm down}) ] } \over
                  { [ N(\ell^+,{\rm up}) +  N(\ell^-,{\rm up}) ]
                + [ N(\ell^+,{\rm down}) + N(\ell^-,{\rm down}) ] } }~.
\end{eqnarray}
This asymmetry does not require final state interactions.
The complex structure needed for this CP-violating asymmetry
comes from the azimuthal phase in the decay process.
Some additional triple product correlations
for top quark pair events have been
studied in Refs.~\cite{treecp,bnos} and \cite{bmm} for $e^-e^+$ and
photon-photon collisions, respectively.

To illustrate the size of CP-violating effects predicted in various models,
we consider a model by Weinberg~\cite{weinberg}.
In this model, the mass matrix of Higgs bosons mixes CP even and odd scalars.
The phenomenological form of the Yukawa interactions
in this model is~\cite{hzero}
\begin{eqnarray}
{\cal L} & = & -{m_t \over v} {\bar t} ( a L + a^* R) t ~,
\end{eqnarray}
where $L$ (or $R$) denotes the left-handed (or right-handed)
projection operator ${1 \over 2} (1-\gamma_5)$
(or ${1 \over 2} (1+\gamma_5)$), and
$v=(\sqrt{2}G_F)^{-1/2} \sim 246\,$GeV. The CP-violating
effect in ${\cal A}_E(\ell)$ is proportional to
${\rm Im}(a^2)=2 {\rm Im}(a){\rm Re}(a)$.
$a$ is a combination of model-dependent mixing angles.
Weinberg showed that for a reasonable choice of Higgs
vacuum expectation values,
$|{\rm Im}(a^2)| \leq \sqrt{2}$. Consequently, ${\cal A}_E(\ell)$
is of the order $10^{-3}$ for $m_t \sim 150\,$GeV and $m_H \sim 100\,$GeV
at the LHC or at the
NLC (Next Linear Collider, an $e^-e^+$ collider with
$\sqrt{S}=500\,$GeV).

To ask how many $t \bar t$ pairs are needed to measure the CP-violating
effects of this order after taking into account the detection
efficiencies, we have performed a study for the decay mode
$t {\bar t} \ra \ell^+ \ell^- \nu \bar{\nu} b {\bar b}$ in
Ref.~\cite{eecp}. We concluded that about $10^7$ $t \bar t$ pairs
are required in electron collisions.
Thus, for a $\sqrt{S}=500\,$GeV $e^-e^+$ collider,
an integrated luminosity of about $10^4 - 10^5$ ${\rm fb}^{-1}$
has to be delivered. This luminosity is
at least a factor of $100$ higher than the
planned NLC colliders.

\subsection{ At Hadron Colliders }
\indent

There have been many studies~\cite{toppol,treecp}
on how to measure the CP-violating effects in the $t \bar t$ system
produced at proton-antiproton or proton-proton colliders,
just as those done for $e^-e^+$ colliders.
The CP-odd observables are similar to what we have
discussed in the previous section.
As shown in Ref.~\cite{brma}, a couple of other examples are
the CP-odd observables
$[\hat{p}_{\rm p} \cdot (\vec{p}_{\ell^+} \times \vec{p}_{\ell^-})]~$
and
$[\hat{p}_{\rm p} \cdot (\vec{p}_{\ell^+} \times \vec{p}_{\ell^-})]
\,
[\hat{p}_{\rm p} \cdot (\vec{p}_{\ell^+} - \vec{p}_{\ell^-})]$,
where $\hat{p_{\rm p}}$ is the direction of motion of the proton
in the CM frame of the pp or $\ppbar$ collision,
$\vec{p}_{\ell^+}$ and $\vec{p}_{\ell^-}$ are the three-momenta of
$\ell^+$ and $\ell^-$ in
$t {\bar t} \ra \ell^+ \ell^- \nu \bar{\nu} b {\bar b}$,
respectively.
We note that although the initial state in a pp collision
(such as at the LHC)
is not an eigenstate of a CP transformation,
these CP-odd observables can still be defined as long as
the production mechanism is dominated by $gg$ fusion. This is
indeed the case for $t \bar t$ pair productions at the LHC.

At hadron colliders, the number of
$t \bar t$ events needed to
measure a CP-violating effect of the order of $10^{-3} - 10^{-2}$
is about $10^7 - 10^8$.
To examine the potential of various current and future
hadron colliders in measuring the CP-violating asymmetries,
let's estimate the total event rates of $t \bar t$ pairs
for a 180 GeV SM top quark produced at these colliders.
At the Tevatron, Di-TeV, and LHC, an integrated luminosity of
10, 100, and 100 ${\rm fb}^{-1}$ will produce about
$4.5 \times 10^4$, $2.6 \times 10^6$, and $4.3 \times 10^7$
$t \bar t$ pairs, respectively \cite{doug}.

\section{ CP violation in Single-Top Productions }
\indent

As discussed in section 3, the top quark can also be produced via
the $W$-gluon fusion process to yield a single-$t$ or single-$\bar t$ event.
In the SM, the top quark  produced by this mechanism is about
one hundred percent
left-handed (longitudinally) polarized~\cite{toppol}.
Given a polarized top quark, one can use
the triple product correlation, as defined in Eq.~(2),
to detect CP violation of the top quark.

For  a polarized top quark, one can either use
$\vec{\sigma}_t \times \vec{p}_b$
or $\vec{p}^{\rm Lab}_t \times \vec{p}_b$
to define the decay plane of $t \ra b W (\ra \ell^+ \nu) $.
Obviously, the latter one is easier to implement experimentally.
Define the asymmetry to be
\begin{eqnarray}
{\cal A}_{io} & \equiv &
      { {   N(\ell^+ \, {\rm out \, of \, the \, decay \, plane})
         -  N(\ell^+ \, {\rm into \, the \, decay \, plane }) } \over
           {   N(\ell^+ \, {\rm out \, of \, the \, decay \, plane})
           +  N(\ell^+ \, {\rm into \, the \, decay \, plane }) } }~.
\end{eqnarray}
If ${\cal A}_{io}$ is not zero, then the time-reversal T is not conserved,
therefore CP is violated for a CPT invariant theory.
Due to the missing momentum of the neutrino from the decay of the $W$-boson,
it is difficult to reconstruct the azimuthal angle ($\phi_W$)
of the $W$-boson from the decay of the top quark. Once the angle
$\phi_W$ is integrated over, the transverse polarization of the top
quark averages out, and only the longitudinal polarization of the top
quark contributes to the asymmetry ${\cal A}_{io}$.
Thus, the asymmetry ${\cal A}_{io}$ can be used to study the
effects of CP violation in the top quark,
which is about one hundred percent left-handed (longitudinally) polarized
as produced from the $W$-gluon fusion process.
To apply the CP-violating observable
${\cal A}_{io}$, one needs to reconstruct the directions of both the $t$
and $b$ quarks.
It has been shown in Ref.~\cite{onetcp} that it takes about
$10^7-10^8$ single-top events
to detect CP violation at the order of $\sim 10^{-3} - 10^{-2}$.

For $m_t=180\,$GeV at the Tevatron, Di-TeV, and LHC,
an integrated luminosity of
10, 100, and 100 ${\rm fb}^{-1}$ will produce about
$2 \times 10^4$, $1.4 \times 10^6$, and $2 \times 10^7$
single-$t$ or single-$\bar t$ events, respectively \cite{doug}.
At the NLC, the single top quark production rate is
much smaller. For a $2\,$TeV electron collider,
the cross sections for
$e^-e^+ \ra e^- {\bar \nu_e} t {\bar b}$
and
$e^+ \gamma \ra {\bar \nu}_e t {\bar b}$
are 8 fb and 60 fb, respectively \cite{eeonetop}.
Hence, it will be extremely difficult
to detect CP violation effects at the order of $\leq 10^{-2}$
in the single-top events produced in electron collisions.

A few comments are in order. First, to extract the {\it genuine}
CP-violating effects, we need to study the difference in the asymmetry
${\cal A}_{io}$ measured in the single-$t$ and single-$\bar t$
events because the time-reversal violation in
${\cal A}_{io}$ of the $t$ (or $\bar t$) alone could be
generated by final state interactions without CP-violating phases.
Second, the detection efficiency for this method is not close
to one, so a good understanding of the kinematics of the decay products and
how the detector works are needed to make this method useful.

The asymmetry ${\cal A}_{io}$ belongs to the second category of
CP-violating observables, and is CP-odd and CP$\hatt$-even. Next, let's
consider another asymmetry ${\cal A}_t$ which belongs to the first
category of CP-violating observables, and is CP-odd and CP$\hatt$-odd.

Another method for detecting CP-violating effects is to make use of
the fact that $\ppbar$ is a CP eigenstate; therefore, the difference in the
production rates for $\ppbar \ra t X$ and $\ppbar \ra \bar t X$
is a signal of CP violation.
This asymmetry is defined to be
\begin{eqnarray}
{\cal A}_t & \equiv &
    { \sigma(\ppbar \ra t X) - \sigma(\ppbar \ra \bar t X) \over
               \sigma(\ppbar \ra t X) + \sigma(\ppbar \ra \bar t X) } ~~.
\end{eqnarray}
As noted in Ref.~\cite{toppol},
the production rate of $\ppbar \ra t X$ is proportional to
the decay rate of $t \ra b W^+$, and the rate of
$\ppbar \ra \bar t X$ is proportional to
the rate of $\bar t \ra \bar b W^-$.
This implies that ${\cal A}_t = {\cal A}_{bW}$, c.f. Eq.~(1).
There have been quite a few models studied in the literature about the
asymmetry in ${\cal A}_{bW}$. For instance, in the Supersymmetric
Standard Model where a CP-violating phase may occur in the
left-handed and right-handed top-squark, ${\cal A}_{bW}$ can
be as large as a few percent
depending on the details of the parameters in the model \cite{stopcp}.

Before we conclude this section, we note that
QCD has the exact symmetries of
C and P, thus ${\cal A}_t$
will not be affected by QCD radiative corrections.
In the next section, we would like to consider a simplified model to
illustrate the possibility of having a large CP-violating asymmetry
${\cal A}_t$.

\section{A Model}
\indent

Consider the CP-violating asymmetry ${\cal A}_t$
in an effective lagrangian containing a neutral Higgs boson $H$ with a
FCNC interaction
\begin{eqnarray}
{\cal L} &  = & {g \over \sqrt{2}} W^-_{\mu} {\bar b} \gamma^{\mu} L t
           +{g \over \sqrt{2}} W^+_{\mu} {\bar t} \gamma^{\mu} L b
\nonumber \\
        & & -{m_t \over v} {\bar t} (a L + a^{*} R) t H
\nonumber \\
        &  & -{\sqrt{m_{t'} m_t} \over v} {\bar t'} (f L + g R) t H
     -{\sqrt{m_{t'} m_t} \over v} {\bar t} (g^{*} L + f^{*} R) t' H ~~,
\end{eqnarray}
where $t'$ is an $SU(2)_L$ singlet field.
As discussed just below Eq.~(5), the existence of a
non-vanishing complex number $a$ in (8) signals CP violation
in the interactions of $t$-$\bar t$-$H$, which however will not
contribute to ${\cal A}_t$ as defined in Eq.~(7) at the one loop level.
To have a non-vanishing
asymmetry ${\cal A}_t$ from this model, a few conditions are required.
First, the model has to have complex couplings, namely, $f$
or $g$ is complex.
Second, the model has to have some other decay modes for the
top quark to generate
the absorptive part of the decay amplitude $t \ra b W^+$.
This is possible if the FCNC decay channel $t \ra t' H$ is allowed
in addition to the tree level decay process $t \ra  b W^+$.
This implies $m_t > m_{t'} + m_H$.
Given the model of (8), it is straightforward to calculate the
asymmetry ${\cal A}_t$.

Denote the one-loop self-energy of the top
quark, with momentum $p$, as
\begin{eqnarray}
-\Sigma(p) & = & 2 A(p) \pslash L + 2 B(p) \pslash R
             +m_t C(p) L + m_t D(p) R~~.
\end{eqnarray}
For $m_{t'} < m_t - m_H$, it is easy to show that \cite{jliu}
\begin{eqnarray}
{\cal A}_t & = &  - {\rm Im}[C(p)]
\nonumber \\
 & = & {1 \over 16 \pi v^2} \left({m_{t'} \over m_t}\right)^2
   \sqrt{\lambda(m_t,m_H,m_{t'})} \, {\rm Im}[ f g^* ] ~~,
\end{eqnarray}
with
\begin{eqnarray}
 \lambda(m_t,m_H,m_{t'}) &  =  &
   \left[ m_t^2- (m_H + m_{t'})^2 \right]
   \left[ m_t^2- (m_H - m_{t'})^2 \right] ~~.
\end{eqnarray}
If ${\rm Im}[ f g^* ] \neq 0$, then CP is violated.
Experimentally, the couplings $f$ and $g$ of $t$-$\bar{t'}$-$H$
interaction are not yet constrained.

To estimate the numerical size of this CP-violating effect, we
need to input values of the mass parameters  $m_t$, $m_H$,
and $m_{t'}$, along with the coupling constants $f$ and $g$.
For the sake of argument, we assume that $m_t= 175\,$GeV,
and $m_H=65\,$GeV.
{}From LEP and SLAC data, the mass of $t'$ can be
as low as $M_Z/2$ assuming a SM coupling of
$t'$-$\bar{t'}$-$Z$. Its mass can be smaller if the coupling of
$t'$-$\bar{t'}$-$Z$ is weaker.
We would argue in the following that $m_{t'}$ can be as large as
90 GeV without deviating from the current collider data.
Consider the direct production of $t'$ at the Tevatron through
the QCD processes $q {\bar q}, gg \ra t' \bar t'$.
The question is: ``What's the lower bound on $m_{t'}$ from direct
search?'' Let's assume that the strong interaction
property of the $t'$ quark is the same as for the
other quarks (such as the top quark) and described by QCD theory.
What has been measured by the experimentalists is the product of
\begin{eqnarray}
\sigma(p \bar p \ra t' {\bar t'}) \cdot
{\rm Br}(t' \ra bW^+ \ra b \ell^+ \nu \, {\rm or} \, b j j)
\cdot
{\rm Br}(\bar{t'} \ra \bar{b} W^- \ra \bar{b} \ell^- {\bar \nu}
\, {\rm or} \, \bar{b} j j)~~.
\end{eqnarray}
In our model, Eq.~(8), $t'$ is an $SU(2)_L$ singlet field; therefore,
it will not directly couple to $b$ and $W^+$. It has to first mix with
the $t$ quark, then decay to $bW$, {\it i.e.},
$ t' \ra t \ra b W$.
Let's assume that this mixing is small, and
$ t' \ra t \ra b W$ is not its dominant decay mode,
so ${\rm Br}(t' \ra bW^+)$ is small.
Then, the lower mass bound for a SM top quark
given by the Tevatron data would not apply to $t'$.
In such a case, $m_{t'}=90\,$GeV would still be possible.
The values of $f$ and $g$ depend on the detail of the models.
Inspired by models with multi-Higgs doublets without the natural flavor
conservation condition \cite{nfc},
$f$ and $g$ can be of the same order as $a$.
For simplicity, let's assume that $f=g^*=\xi a$, where $\xi$ is
a real number of ${\cal O}(1)$.
We note that in Eq.~(8) we have defined the coupling constants
$a$, $f$ and $g$ such that the $t$-$\bar t$-$H$ coupling is
proportional to $m_t/v=\sqrt{m_t^2}/v$ and the
$t$-$\bar{t'}$-$H$ coupling is proportional to
$\sqrt{m_{t'} m_t}/v$. By doing so, one has assumed that the
interactions of the Higgs boson to fermions are related to
how the masses of the fermions were generated \cite{nfc,hou}.
After substituting the above parameters in Eq.~(10), we obtain
\begin{eqnarray}
{\cal A}_t & = & 1.2 \, \xi^2 \, {\rm Im}(a^2) \times 10^{-3} ~~.
\end{eqnarray}
For $\xi \sim 3$, ${\cal A}_t$ can be as large as a few percent
for $|{\rm Im}(a^2)| < \sqrt{2}$.

Next, let's examine how many top quark events are needed to
detect a few percent effect in the CP-violating asymmetry ${\cal A}_t$.
Consider $t \ra b W^+ \ra b \ell^+ \nu$, where $\ell= e \, {\rm or} \,
\mu$. Its branching ratio $B_W$ is calculated by the product of
${\rm Br}(t \ra bW^+)$ and  ${\rm Br}(W^+ \ra \ell^+ \nu)$, where
${\rm Br}(W^+ \ra \ell^+ \nu)$ is $2/9$ and
\begin{eqnarray}
{\rm Br}(t \ra bW^+) & =&  { \Gamma(t \ra b W^+) \over
        \Gamma(t \ra b W^+) + \Gamma(t \ra t' H) }~~,
\nonumber \\
\Gamma(t \ra b W^+) &  =&  {1 \over 8 \pi v^2} m_t M_W^2
   \left(1-{M_W^2 \over m_t^2} \right)^2 \left( 1 + {m_t^2 \over 2 M_W^2}
  \right) ~~,
\nonumber \\
\Gamma(t \ra t' H)  & = &   {1 \over 16 \pi v^2}
    \sqrt{\lambda(m_t,m_H,m_{t'})}
\nonumber \\
& & \cdot
    {m_{t'} \over m_t^2} \left[ (|f|^2+|g|^2) (m_t^2+m_{t'}^2-m_H^2)
           +4 {\rm Re}(f g^*) m_t m_{t'} \right]~~.
\end{eqnarray}
In addition to the assumption that $f=g^*=\xi a$, we further assume
${\rm Re}(a)={\rm Im}(a)$ for simplicity, so $|a|^2={\rm Im}(a^2)~$
and
\begin{eqnarray}
{\rm Br}(t \ra b W^+) & = & {1.56 \over {1.56 + 1.8 \, \xi^2 \,
{\rm Im}(a^2) } }~~,
\end{eqnarray}
which is about $0.38$ (or $0.17$) for $\xi=1$ (or 3)
after taking ${\rm Im}(a^2)=\sqrt{2}$.
(In the SM, ${\rm Br}(t \ra b W^+) \approx 1$.)
Hence, we have $B_W=0.38 \times 2/9=0.084$ for $\xi=1$ and $0.038$ for
$\xi=3$.

Let's assume that the efficiency of $b$-tagging ($\epsilon_b$)
is about 15\%, and the kinematic acceptance
($\epsilon_k$) of reconstructing the single-top event,
$\ppbar \ra t X \ra b W^+ X \ra b \ell^+ \nu X$,
is about 50\% from a Monte Carlo study ~\cite{wgone,doug}.
The number of single-$t$ and single-${\bar t}$
events needed to measure ${\cal A}_t$ is
\begin{eqnarray}
{\cal N}_t & = & { 1 \over B_W \epsilon_b \epsilon_k}
  \left(1 \over {\cal A}_t \right)^2 ~~.
\end{eqnarray}
Thus, to measure ${\cal A}_t$ of a few percent,
${\cal N}_t$ has to be as large as $\sim 10^6$, which corresponds
to an integrated luminosity of 100 $\ifb$ at the Di-TeV.

\section{ Conclusions }
\indent

I have discussed the strategies for measuring the CP asymmetries
in the top quark system, either in $t \bar t$ pair events
or in single-$t$ or single-$\bar t$ events,  for
the $e^-e^+$, $e^+ \gamma$, pp and $\ppbar$ colliders.
In general, models with a CP-violating mechanism predict
CP asymmetries of the order of $10^{-3} - 10^{-2}$ percent,
which will require, after unfolding the detection efficiencies,
about $10^7 - 10^8$ top quark events to be produced.
Therefore, we conclude that it would be difficult to measure
a CP-violating asymmetry smaller than $10^{-3}$ even at the LHC.
G. Kane drew the same conclusion from examining how well the
detectors can measure
a CP-odd observable in colliders \cite{imkane}.
If the transverse polarization asymmetry from the
$t \ra b W^+ (\ra \tau^+ \nu_\tau)$ decay can be measured, then
$\sim 10^2$ fewer top quark pairs are needed \cite{taupol}.
However, it would be a bigger challenge to measure this asymmetry
experimentally in hadron collisions than in electron collisions.
For a large CP-violating asymmetry in single-top
events, a $\ppbar$ collider (such as the Tevatron)
offers an unique opportunity for measuring the asymmetry
${\cal A}_t$ by simply counting the
difference in the single-$t$ and the single-$\bar t$
production rates.

\section{ Acknowledgments }
\indent

I thank G. Kane  and G.A. Ladinsky for collaboration, to them and
R. Brock, Darwin Chang and Wei-Shu Hou, for helpful discussions.
I also thank D.O. Carlson and G.A. Ladinsky
for a critical reading of the manuscript.
This work was supported in part by an NSF grant No. PHY-9309902.

\newpage
\def\jourtpol#1&#2{{\it #1}{\bf #2}}
\def\journal#1&#2(#3)#4{
{\unskip,~\it #1\unskip~\bf\ignorespaces #2\unskip~\rm (19#3) #4}}

\newpage

\end{document}
\end